\documentclass[reprint,
 amsmath,amssymb,
 aps,
]{revtex4-2}

\usepackage{graphicx}
\usepackage{dcolumn}
\usepackage{bm}
\usepackage{url}

\begin{document}

\preprint{APS/123-QED}

\title{Comments on ``Direct measurement of the ionization quenching factor of nuclear recoils in germanium in the keV energy range"}

\author{J.I. Collar}
\email{collar@uchicago.edu}
\author{C.M. Lewis}%
\affiliation{%
Enrico Fermi Institute, Kavli Institute for Cosmological Physics, and Department of Physics\\
University of Chicago, Chicago, Illinois 60637, USA
}%


\date{\today}

\begin{abstract}
We examine a recent measurement of the quenching factor (QF) in germanium at 80 K, noticing a number of inconsistencies capable of  affecting a claimed agreement with Lindhard's ion-stopping formalism in the sub-keV nuclear recoil energy regime. Namely,  an underestimated uncertainty in the energy scale and a missing correction for a large instrumental non-linearity in this scale able to severely distort the QF behavior at low energy, favoring reduced values.  The discussion is expanded to inspect the impact of QF model selection on a study of neutrino electromagnetic properties using CE$\nu$NS data that supports a non-zero electric charge at the 3.5 $\sigma$ level. 
\end{abstract}

\maketitle


Monoenergetic neutron beams are commonly used to produce low-energy nuclear recoils (NRs) like those expected from coherent elastic neutrino-nucleus scattering (CE$\nu$NS) and  some dark matter  candidates. A recent study of this type \cite{conusqf} has measured the response to NRs in germanium diodes operated at liquid-nitrogen temperature, claiming a good agreement with Lindhard's theory of ion energy loss \cite{lindhard} down to 0.4 keV$_{nr}$ (the suffix ``nr" stands for nuclear recoil, ``ee" for electron-equivalent, or ionization energies). This response is quantified by a quenching factor (QF) equal to the ratio of ionization yields produced by NRs and electron recoils of the same energy. The behavior of the sub-keV$_{nr}$ QF critically impacts the interpretation of CE$\nu$NS measurements \cite{csiqf,qf,ncc}.

The device described in \cite{conusqf} consists of four independent pixels, each acting as a separate detector. The systematic uncertainty  in the energy scale (repeatedly claimed to be a ``conservative" 10 eV$_{ee}$) is determined by the uncertainty in the position of a tiny peak arising from neutron activation, expected at $\sim$1.297 keV$_{ee}$. Spectra from all four pixels are co-added in order to obtain barely sufficient statistics for a fit (Fig.\ 4 in \cite{conusqf}). The authors point to a lack of impact from  choice of background model or bin width on the quality of their fit. Noticing the unnaturally-narrow energy region used for this fit, the resulting tension between the best-fit sigmoid background model \footnote{No information on this step-like feature is offered. This  is uncharacteristic of HPGe spectra, meriting discussion. } and  neighboring data, and an absence of mention about the effect of fitting window choice, we explored the latter. While we can reproduce a 10 eV$_{ee}$ uncertainty in peak position, for arbitrary fitting windows selected within 0.5-2.5 keV$_{ee}$ the best-fit centroid returned is scattered over a broad energy range, \mbox{21 eV$_{ee}$} wide. This invalidates the professed approach to defining the energy uncertainty. 

Prompted by this test, we carefully examined other assertions made in \cite{conusqf}. The energy scale is defined by  two neighboring $^{55}$Fe x-rays at 5.90 and 6.49 keV$_{ee}$, measured with high statistics (Fig.\ 3 in \cite{conusqf}). In addition to these, two neutron-activation peaks also in close vicinity to each other at 9.87 and 10.37 keV$_{ee}$ are used to provide a linear fit to the energy. These are presented as having ``sufficient statistics" for this purpose. From the merged (all data, all pixels) spectra in Fig.\ 4 of \cite{conusqf} we derive that a total of $\sim$450 and $\sim$1,100 counts were available under these peaks, respectively. The inset of Fig.\ 1 below shows the typical calibration data quality in a best-case scenario (when the total available statistics is divided by four) under the conservative assumption that a single energy calibration was required per pixel.

\begin{figure}[!htbp]
\includegraphics[width=.82 \linewidth]{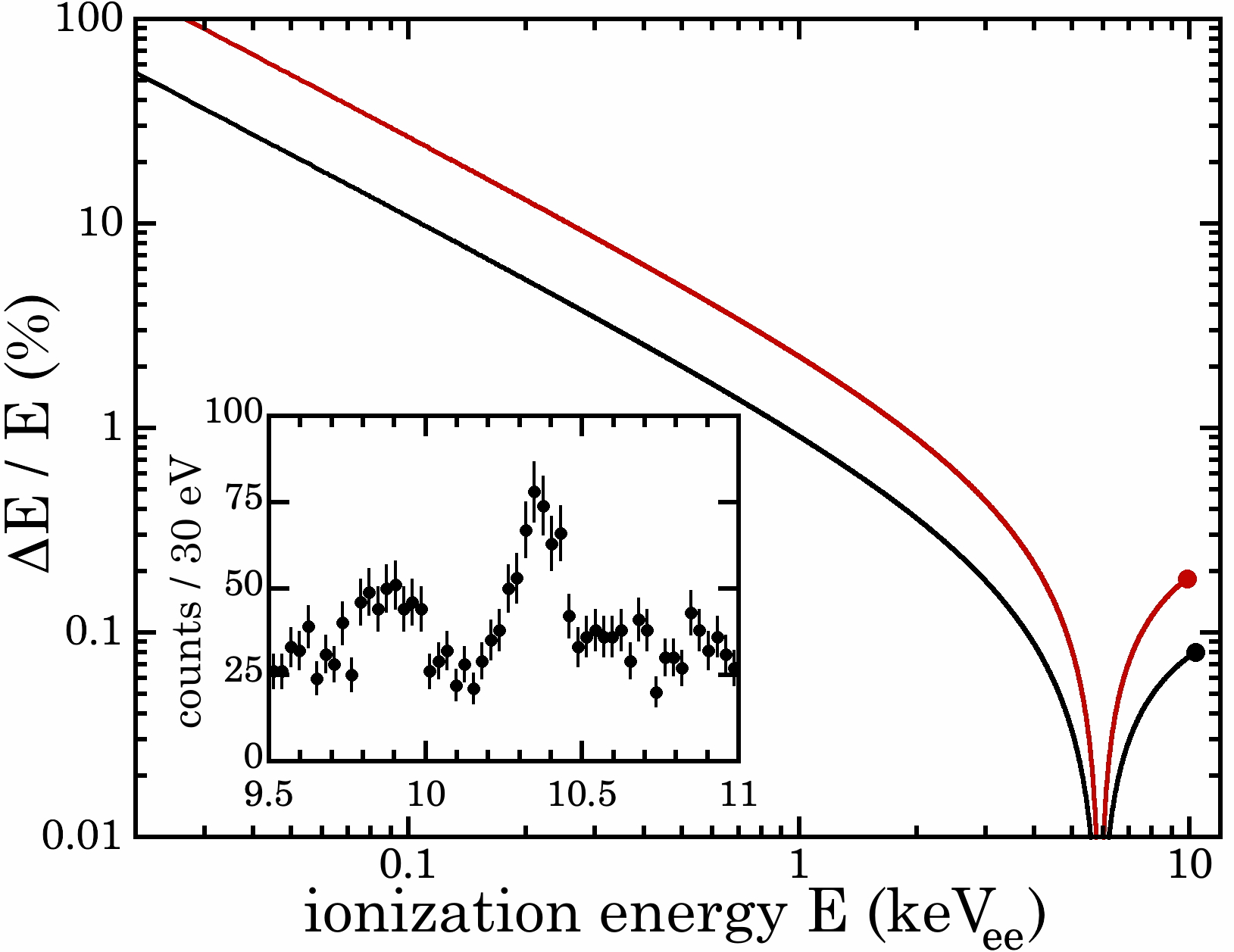}
\caption{\label{fig:epsart} Lower limit to the fractional energy uncertainty expected from a linear fit using the 9.87 keV$_{ee}$ (red) or 10.37 keV$_{ee}$ (black) calibration peaks, assuming a perfectly-defined fulcrum at 5.9 keV$_{ee}$ (see text). The inset shows a representative simulated example of maximum calibration statistics available per detector pixel for these two peaks in \cite{conusqf}.}
\end{figure}

This is visibly insufficient and reminiscent of the proverbial ``two-point linear fit" energy calibration where a large uncertainty in the reconstructed position of a high-energy peak is  amplified at low energy. Fig.\ 1 illustrates this well-known effect, using peak position uncertainties derived from an ensemble of 10$^{4}$ Monte Carlo-generated data sets like those in the inset, fitted by two Gaussians and a flat background. Extending this simulation to include the $^{55}$Fe peaks, assumed to be both defined with modest 3 eV$_{ee}$ uncertainty \footnote{Unable to digitize Fig.\ 3a in \cite{conusqf}, we use a 0.05\% stability from Fig.\ 3b for the most intense of these $^{55}$Fe peaks  to conservatively assess the uncertainty in their positions.}, we obtain a dispersion (\frenchspacing{r.m.s.}) in the offset (independent term) of the corresponding four-point linear fits of $\pm$16.2 eV$_{ee}$. Sec.\ 2.3 in \cite{conusqf} describes the magnitude of these offsets, extracted from an undetermined (possibly larger than four) number of energy calibrations, as ``typically varying" between -80 and 25 ADC bins. Using the 2,600 ADC bins/keV$_{ee}$ slope quoted, this corresponds to a 40 eV$_{ee}$ ``typical" range for the energy calibration offsets, confirming our seemingly conservative simulated estimate. Identifying the  spread in these offsets with a measure of the uncertainty in the low-energy scale leads us to conclude that it has been significantly underestimated at 10 eV$_{ee}$ by the authors of \cite{conusqf}. For perspective, the iron-filter sub-keV$_{nr}$ QF measurements in \cite{qf} used alpha-induced x-ray emission to generate six high-statistics \footnote{Specifically, 2,550 counts per peak on average.} calibration peaks spread over the 4-7 keV$_{ee}$ range. This resulted in excellent fit linearity  (Pearson's R = 0.99998), and yet an energy uncertainty of 11 eV$_{ee}$ \cite{qf}. 

We expect that a more balanced appraisal of the energy uncertainty will generate an increase in the correlated systematic uncertainty band in Fig.\ 14 of \cite{conusqf}, better allowing for deviations from Lindhard theory at low energy like those noticed in \cite{qf}. However, this brings us to our main concern. Sec.\ 2.3 in \cite{conusqf} introduces two sources of observed non-linearity in the energy scale. The first is positive, i.e., in the direction of making deposited energies appear larger than their actual values. It corresponds to the commonplace effect of the signal acceptance curve, which is properly accounted for in \cite{conusqf} as well as in most every other study of this type. In this sense, to describe it as a non-linearity can be regarded as a bit of a misnomer. 

\begin{figure}[!htbp]
\includegraphics[width=.8 \linewidth]{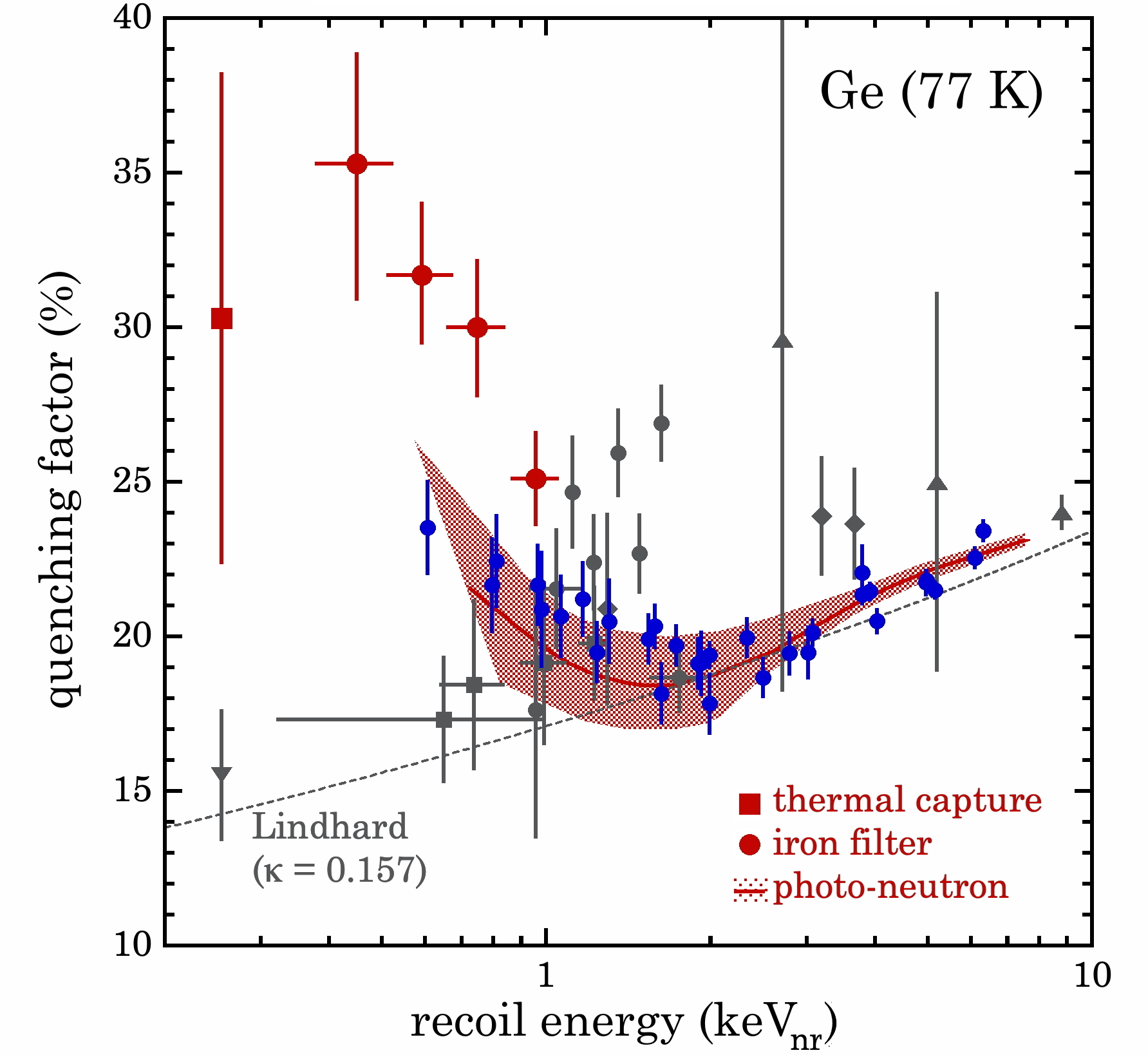}
\caption{\label{fig:epsart} Estimated impact of a correction for negative  non-linearity in the energy scale absent from \cite{conusqf}. Blue points are derived by extracting an average ionization energy from  QF and NR energies in Fig.\ 14 of \cite{conusqf}, applying the correction as approximated in this note, and recomputing the QF. Red data are from \cite{qf}.}
\end{figure}

The second source of non-linearity is negative, and specific to the data-acquisition system utilized in \cite{conusqf}. From a brief description provided, it originates in a ballistic deficit in energy determination from use of a short shaping time in combination with a delay in the trigger time stamp, progressively larger towards low signal energy. It is quantified as follows: ``less than 5 eV$_{ee}$ at 400 eV$_{ee}$ and reaches about 25 eV$_{ee}$ at 250 eV$_{ee}$". This last energy corresponds to the ionization  deposited by a $\sim$1.3 keV$_{nr}$ NR obeying  Lindhard's theory. No further information is provided. This is surprising, given that the analysis in \cite{conusqf} involves data points as low as 25 eV$_{ee}$, and Fig.\ 9 in \cite{conusqf} confirms that this instrumental effect should continue to grow in magnitude as signal energy decreases. Assuming for the sake of argument that the trend described  is maintained, this non-linearity would be a sobering $\sim$50\% at 100 eV$_{ee}$, with the net result of making the sub-keV$_{nr}$ QF appear artificially smaller by as much. Inspection of Fig.\ 13 in \cite{conusqf} confirms that a correction for this effect, had it been implemented, could generate QF values of order $\sim$28\% at 0.4 keV$_{nr}$, i.e., a deviation from Lindhard very similar to that found in \cite{qf}. Fig.\ 2 in this note further illustrates this point.

Making this correction seems possible, seeing that this negative non-linearity can be quantified. It is nowhere applied. Instead, a fraction of Sec.\ 4.1 in \cite{conusqf} is dedicated to the argument that allowing an energy shift by the claimed 10 eV$_{ee}$ uncertainty is sufficient to account for this effect (as per the extrapolation above, this is clearly not the case). Specifically: ``The present approach is therefore conservative and also covers uncertainties related to the small non-linearities observed in the few hundreds of eV$_{ee}$ region". While this is true for QF results obtained well above 1 keV$_{nr}$, we find the present treatment of lower NR energies incautious and lacking. The authors conclude by  making the point that CONUS HPGe detectors are not  sensitive yet to sub-keV$_{nr}$ QF specifics, in view of their relatively high energy threshold. Other germanium CE$\nu$NS searches do not share this limitation \cite{ncc}, begging for a more careful analysis. 

We have recently emphasized \cite{csiqf,qf,ncc} the importance of a robust knowledge of low-energy quenching factors, for CE$\nu$NS to reach its full potential in the search for new physics. A recent analysis of CsI[Na] CE$\nu$NS data inadvertently but beautifully illustrates this point. This work \cite{amir} finds a preference as large as 3.5 $\sigma$ for neutrino electromagnetic properties (finite millicharge and/or a non-zero magnetic moment) at levels already severely constrained by other searches. It nevertheless adopts a QF model \cite{cqf}  prejudiced to favor a subset of experimental determinations, disregarding other available QF measurements. The smaller Standard Model CE$\nu$NS signal rate prediction generated by an unbiased, physics-based CsI[Na] QF model \cite{csiqf} should obviate the need to involve exotic neutrino properties, as of today. Arguably, CE$\nu$NS data are already informing the relative quality of QF models.

We encourage the authors of \cite{conusqf} to revisit their treatment of  sub-keV$_{nr}$ QF data. On our side we plan to implement the method described in \cite{french} to test the Lindhard formalism for HPGe at 0.4 keV$_{nr}$, using once more the high-purity thermal neutron beam at OSURR \cite{qf,ohio}. 

\bibliography{apssamp}

\end{document}